\documentstyle[floats,prd,aps,epsfig,eqsecnum,12pt]{revtex}
\makeatletter
\newbox\tempboxa
\newdimen\captionboxsubcount
\def\capsize#1{\captionboxsubcount=#1pt}
\newdimen\captionboxsub
\captionboxsub=\hsize \advance\captionboxsub by -\captionboxsubcount
\advance\captionboxsub by -\captionboxsubcount
\long\def\@makecaption#1#2{
 \setbox\@tempboxa\hbox{#1 #2}
 \ifdim \wd\@tempboxa >\captionboxsub
\rightskip=\captionboxsubcount \leftskip=\captionboxsubcount #1 #2
\else \hbox to\hsize{\hfil\box\@tempboxa\hfil}
 \fi}
\makeatother
\capsize{30}

\begin{document}
\begin{titlepage}
\begin{flushright}
\begin{minipage}{5cm}
\begin{flushleft}
\small
\baselineskip = 10pt
YCTP-P-28-99\\ hep-ph/yymmmnn\\
\end{flushleft}
\end{minipage}
\end{flushright}
\begin{center}
\Large\bf
Enhanced Global Symmetries and a\\
       Strong Electroweak Sector
\end{center}
\footnotesep = 12pt
\vfill
\begin{center}
\large
\centerline{Francesco {\sc Sannino} \footnote{Electronic address : {\tt
francesco.sannino@yale.edu}}}
\vskip 1cm
 {\it Department of Physics, Yale University, New
Haven,\\~CT~06520-8120,~USA.}
\end{center}
\vfill
\begin{center}
\bf Abstract
\end{center}
\begin{abstract}
\baselineskip = 17pt
In this talk I review the intriguing possibility Ref.~\cite{ARS}
that the physical spectrum of a vector-like gauge field theory
exhibits an enhanced global symmetry near a chiral phase
transition. A transition from the Goldstone phase to the symmetric
phase is expected as the number of fermions $N_f$ is increased to
some critical value. Various investigations have suggested that a
parity-doubled spectrum develops as the critical value is
approached. Using an effective Lagrangian as a guide one observes
that a parity doubling is associated with the appearance of an
enhanced global symmetry in the spectrum of the theory. If such a
near-critical theory describes symmetry breaking in the electroweak
theory, the additional symmetry suppresses the contribution of the
parity doubled sector to the $S$ parameter.
\end{abstract}
\vfill
\end{titlepage}

\section{Introduction}

Gauge field theories exhibit many different patterns of infrared
behavior. Indeed for a vector-like theory such as QCD, it is known
that for low values of $N_f$, the theory confines and chiral
symmetry breaking occurs. On the other hand, for large $N_f$ the
theory loses asymptotic freedom. In between, there is a conformal
window where the theory does not confine, chiral symmetry is
restored, and the theory acquires a long range conformal symmetry.
It has been proposed that for an $SU(N)$ gauge theory, there is a
transition from the confining, chirally broken theory to the
chirally symmetric theory at $N_f\approx 4 N$~\cite{ATW,SS}. Recent
lattice simulations, however, seem to indicate ~\cite{mawhinney}
that the amount of chiral symmetry breaking decreases substantially
(for $N=3$) when $N_f$ is only about $4$.

Assuming that a single transition takes place at some critical
value of $N_f$, we can ask questions about the spectrum of the
theory near the transition. In Reference~{\cite{AS}}, it was argued
by studying Weinberg spectral function sum rules that for
near-critical theories parity partners become more degenerate than
in QCD-like theories. This leads naturally to the idea that parity
doublets might form as chiral symmetry is being restored. Lattice
studies also indicate such a possibility~\cite {mawhinney}.

Here I review the ideas presented in Ref.~\cite{ARS} where it was
observed using an effective-Lagrangian as a guide, that the
formation of degenerate parity partners is associated with the
appearance of an enhanced global symmetry in the spectrum of
states. This new symmetry could play a key role in describing a
possible strong electroweak Higgs sector. Whether the new symmetry
can be shown to emerge dynamically from an underlying gauge theory
with $N_f$ near a critical value remains an open question.

First, in Section \ref{ELSUSU}, the appearance of enhanced global
symmetry is discussed. Confinement is assumed and the symmetry of
the underlying gauge theory, $SU_L(N_f)\times SU_R(N_f)$, is built
into an effective Lagrangian describing the physical states of the
theory. Parity invariance is imposed and the usual pattern of
chiral symmetry breaking ($SU_L(N_f)\times SU_R(N_f)\rightarrow
SU_V(N_f)$) is assumed. The $N_f^2-1$ Goldstone bosons appear
together with scalar chiral partners. We augment the spectrum with
a set of vector fields for both the $SU_L(N_f)$ and $SU_R(N_f)$
symmetry groups. The Lagrangian thus takes the form of a linear
sigma model coupled to vectors.

Analyzing the spectrum one recognizes that there is a particular
choice of the parameters that allows for a degenerate vector and
axial-vector, while enlarging the global symmetry to include an
additional (unbroken) $SU_L(N_f)\times SU_R(N_f)$. This happens as
the spectrum of the theory splits into two sectors with one
displaying the additional symmetry. I then review the arguments
(see Ref.~{\cite{AS}}) that an underlying near-critical $SU(N)$
gauge theory might naturally lead to a more degenerate vector-axial
spectrum than in QCD, and to an enhanced symmetry. Finally we note
that even a discrete additional symmetry, $Z_{2L}\times Z_{2R}$, of
the effective theory is adequate to insure the mass degeneracy of
the vector and axial vector.

The possible appearance of an additional, continuous symmetry was
considered by Casalbuoni et al in Refs.{\cite{sei,sei2}}.

Finally in Section \ref{ELWSU} the electroweak gauge group is
embedded within the global symmetry group. As observed in
Ref.~\cite{ARS} the enhanced symmetry of the strongly interacting
sector, which now provides electroweak symmetry breaking, plays an
important role. The additional symmetry operates as a partial
custodial symmetry for the electroweak $S$ parameter, in the sense
that the parity doubled part of the strong sector, by itself, makes
no contribution to $S$.


 \section{Effective Lagrangian for $SU_L(N_f)\times SU_R(N_f)$ global symmetry}

\label{ELSUSU}

To discuss the possible appearance of enhanced symmetry some
description of the spectrum is needed. It is helpful to use an
effective Lagrangian possessing $SU_L(N_f)\times SU_R(N_f)$
symmetry, the global invariance of the underlying gauge theory.
Chiral symmetry is broken according to the standard pattern $
SU_L(N_f)\times SU_R(N_f)
\rightarrow SU_V(N_f)$. The $N_f^2
-1$ Goldstone bosons are encoded in the $N_f\times N_f$ real traceless
matrix $\Phi^i_j$ with $i,j=1,..., N_f$. The complex matrix
$M=S+i\Phi$ describes both the Goldstone bosons as well as
associated scalar partners $S$. It transforms linearly under a
chiral rotation of the type $M\rightarrow u_L M u^{\dagger}_R$ with
$u_{L/R}$ in $SU_{L/R}(N_f)$.

To augment the massive spectrum one introduces vector and axial
vector fields following a method outlined in Ref.~\cite{joe}. One
first formally gauge the global chiral group introducing the
covariant derivative
\begin{equation} D^{\mu} M =
\partial^{\mu} M - i\tilde{g} A^{\mu}_{L} M + i \tilde{g} M A^{\mu}_R \ ,
\end{equation}
where $A_{L/R}^{\mu}=A_{L/R}^{\mu,a}T^a$ and $T^a$ are the
generators of $SU(N_f)$, with $a=1,...,N_f^2-1$ and $
\displaystyle{{\rm Tr}\left[T^a T^b\right]=\frac{1}{2} \delta^{ab}}$. Under a
chiral transformation
\begin{equation}
  A_{L/R}^{\mu} = u_{L/R} A_L^{\mu} u_{L/R}^{\dagger} -
\frac{i}{\tilde{g}}
\partial^{\mu}u_{L/R} u^{\dagger}_{L/R}.
\label{gaugetrans}
\end{equation}

The effective Lagrangian needs only to be invariant under global
chiral transformations. Including terms only up to mass dimension
four, it may be written in the form
\begin{eqnarray}
L&=&\frac{1}{2}\,{\rm Tr}
\left[D_{\mu}M D^{\mu}M^{\dagger}\right] + m^2 \, {\rm
Tr}\left[A_{L\mu}A_L^{\mu} +A_{R\mu}A_R^{\mu}\right] + h\, {\rm
Tr}\left[A_{L \mu}M A_R^{\mu}M^{\dagger}\right]\nonumber \\
&+&{r}\,{\rm Tr}
\left[A_{L\mu}A_L^{\mu}MM^{\dagger} + A_{R\mu}A_R^{\mu}M^{\dagger}M
\right]  + i\, \frac{s}{2}\,{\rm Tr}\left[A_{L\mu}\left(M
D^{\mu}M^{\dagger}-D^{\mu}M M^{\dagger} \right) \right. \nonumber
\\   &+& \left. A_{R\mu}\left(M^{\dagger}D^{\mu}M -
D^{\mu}M^{\dagger}M\right)\right] \ .
\label{asLagrangian}
\end{eqnarray}
The parameters $h,r$ and $s$ are dimensionless real parameters,
while $m^2$ is a common mass term. To this, we may add a kinetic
term for the vector fields
\begin{equation}
L_{{\rm Kin}}=-\frac{1}{2}\,{\rm Tr}\left[F_{L\mu \nu} F^{\mu
\nu}_L +
F_{R\mu \nu} F^{\mu \nu}_R \right] \ ,  \label{kin}
\end{equation}
 where $F^{\mu \nu}_{L/R}=
\partial^{\mu}A_{{L/R}}^{\nu}-\partial^{\nu} A_{{L/R} }^{\mu} - i
\tilde{g}
\left[A_{{L/R}}^{\mu},A_{{L/R}}^{\nu}\right] \label{F}$
along with vector-interaction terms respecting only the global
symmetry. Finally, one may add the double trace term,
$\displaystyle{{\rm Tr}\left[M M^{\dagger}\right]{\rm
Tr}\left[A_L^2 + A_R^2
\right]} \label{tracedouble}$ at the dimension-four level. To arrange for symmetry breaking, a
potential $V(M,M^{\dagger})$ must be added. When the effective
Lagrangian is extended to the dimension-six level and higher, many
new operators enter. Parity is also a symmetry. It is worth noting
that a Lagrangian of this type has been used  to describe  the
low-lying QCD resonances and interactions \cite{effective
Lagrangians}.

The scalar vacuum expectation value is $v$ and the new vector
fields are
\begin{equation}
V=\frac{A_L + A_R}{\sqrt{2}}\ , \qquad\qquad A=\frac{A_L -
A_R}{\sqrt{2}}
\ .
\label{va}
\end{equation}
The vector-axial vector mass difference is given by
\begin{equation}
M^2_A - M^2_V = v^2\, \left[\tilde{g}^2 - 2 \tilde{g}\,s - h
\right] \ .
\label{differenza}
\end{equation}
In QCD this difference is known experimentally to be positive, a
fact that can be understood by examining the Weinberg spectral
function sum rules (see Ref.~{\cite{BDLW}} and references
therein). The effective Lagrangian description is of course
unrestrictive. Depending on the values of the $\tilde{g}$, $s$
and $h$ parameters, one can have a degenerate or even inverted
mass spectrum.

{\it What kind of underlying gauge theory might provide a
degenerate or inverted spectrum?} Clearly, it has to be different
from QCD, allowing for a modification of the spectral function sum
rules. In Reference~\cite{AS}, an $SU(N)$ gauge theory (with $N>2$)
and $N_f$ flavors was considered.  If $N_f$ is large enough but
below $11N/2$, an infrared fixed point of the gauge coupling
$\alpha_*$ exists, determined by the first two terms in the $\beta$
function. For $N_f$ near $11N/2$, $\alpha_*$ is small and the
global symmetry group remains unbroken. For small $N_f$, on the
other hand, the chiral symmetry group $
\displaystyle{SU_L(N_f)\times SU_R(N_f)}$ breaks to its diagonal subgroup.
One possibility is that the transition out of the broken phase
takes place at a relatively large value of $N_f/N (\approx 4)$,
corresponding to a relatively weak infrared fixed
point~\cite{ATW,SS}. An alternate possibility is that the
transition takes place in the strong coupling regime,
corresponding to a small value of $N_f/N$~\cite {mawhinney}. The
larger value emerges from the renormalization group improved gap
equation, as well as from instanton effects~\cite{ASe}, and
saturates a recently conjectured upper limit~\cite{ACS}. It
corresponds to the perturbative infrared fixed point $\alpha_*$
reaching a certain critical value $\alpha_c$. A similar result
has also been obtained by using a suitable effective Lagrangian
\cite{SS}.

In Reference~{\cite{AS}} the spectrum of states in the broken phase
near a large-$N_f/N$ transition was investigated using the spectral
function sum rules. It was shown that the ordering pattern for
vector-axial hadronic states need not be the same as in QCD-like
theories (small $N_f/N$). The crucial ingredient is that these
theories contain an extended "conformal region" extending from
roughly $2\pi F_{\pi}$ to the scale $\Lambda$ where asymptotic
freedom sets in. In this region, the coupling remains close to an
approximate infrared fixed point and the theory has an approximate
long range conformal symmetry. It was argued that this leads to a
reduced vector-axial mass splitting, compared to QCD-like theories.
This suggests the interesting possibility that parity doublets
begin to form as chiral symmetry is being restored. That is, the
vector-axial mass ratio approaches unity as the masses decrease
relative to $\Lambda$. Lattice results seem to provide supporting
evidence for such a possibility~\cite{mawhinney}, although at
smaller values of $N_f/N$.

If a parity doubled spectrum does appear, it is natural to expect
it to be associated with some new global symmetry. While it is hard
to demonstrated the appearance of a new global symmetry using the
underlying degrees of freedom, one can explore aspects of parity
doubling at the effective Lagrangian level. Returning to this
description, we note that vector-axial parity doubling corresponds
to the parameter choice (see Eq. (\ref{differenza})),
\begin{equation}
\tilde{g}^2 = 2 \tilde{g}\,s + h \ .
\label{degenerate}
\end{equation}
This condition does not yet reveal an additional symmetry and
therefore there is no reason to expect parity degeneracy to be
stable in the presence of quantum corrections and the many higher
dimensional operators that can be added to the effective
Lagrangian in Eq.~(\ref{asLagrangian}).

However, for the special choice $s = \tilde{g}$, $r
=\tilde{g}^2/2$
and $h=-\tilde{g}^2$, the effective Lagrangian acquires a new
continuous global symmetry that protects the vector-axial mass
difference. The effective Lagrangian at the dimension-four level
takes the simple form
\begin{equation} L=\frac{1}{2}\,{\rm Tr}
\left[\partial_{\mu}M \partial^{\mu}M^{\dagger} \right] + m^2 \, {\rm
Tr}\left[A_{L\mu}A_L^{\mu} +A_{R\mu}A_R^{\mu}\right],
  \label{dlag}
\end{equation}
along with vector kinetic and interaction terms, the interaction
term Eq.~(\ref{tracedouble}), and the symmetry breaking potential
$V(M,M^{\dagger})$. The theory now has two sectors, with the vector
and axial vector having their own unbroken global $SU_L(N_f)\times
SU_R(N_f)$. The two sectors interact only through the product of
singlet operators. The full global symmetry is
$\left[SU_L(N_f)\times SU_R(N_f)\right]^2\times U_V(1)$
spontaneously broken to $ SU_V(N_f)\times U_V(1) \times
\left[SU_L(N_f)\times SU_R(N_f)\right]$. The vector and axial
vector become stable due to the emergence of a new conservation
law. This enhanced symmetry would become exact only in the chiral
limit. For finite but small (relative to $\Lambda$) values of the
mass scales in Eq. (\ref{dlag}), there are additional, smaller
terms giving smaller mass splittings and small width-to-mass
ratios.

It is of course a simple observation that a new symmetry and
conservation law emerge if a theory is split into two sectors by
setting certain combinations of parameters to zero. But here we
were led to this possibility by looking for a symmetry basis for
the parity doubling that has been hinted at by analyses of the
underlying gauge theory.

It is worth mentioning that along with the additional global
symmetry $SU_L(N_f)\times SU_R(N_f)$, the effective Lagrangian Eq.
(\ref{dlag}) possesses a discrete $Z_{2L}\times Z_{2R}$ symmetry.
Under $Z_{2L}\times Z_{2R}$ the vector fields transform according
to $A_L \rightarrow z_{L} A_L,~A_R \rightarrow z_{R}A_R$  with
$z_{L/R}=1,-1$ and $z_{L/R}\in Z_{2L/R}$. Actually, the discrete
symmetry alone is enough to insure vector-axial mass degeneracy and
stability against decay. In that case, additional interaction
terms, such as the single trace term $r{\rm Tr} \left[A_{\mu L}
A^{\mu}_L M M^{\dagger} + A_{\mu R} A^{\mu}_R M^{\dagger} M
\right]$ are allowed, but degeneracy and stability are still
insured.

\section{Strongly Interacting Electroweak Sector} \label{ELWSU}

To discuss the consequences of enhanced symmetry for a strong
symmetry breaking sector of the standard electroweak theory one
embeds the $SU_L(2) \times U_Y(1)$ gauge symmetry in the global
$SU_L(N_f)\times SU_R(N_f)$ chiral group. For simplicity I restrict
attention to the $SU_{L}(2)\times SU_R(2) $ subgroup of the full
global group. The electroweak gauge transformation then takes the
form $M \rightarrow u_W M u_Y^{\dagger}$. $M$ is now a $2
\times 2$ matrix which can be written as
$M=\frac{1}{\sqrt{2}} \left[\sigma +i
\vec{\tau}\cdot\vec{\pi}\right]$ , where
$u_W=u_L=\exp\left(\frac{i}{2}\epsilon^a
\tau^a\right)$
with $\tau^a$ the Pauli matrices, and where
$u_Y=\exp\left(\frac{i}{2}\epsilon_0 \tau^3\right)$. The weak
vector boson fields transform as
\begin{equation}
W^{\mu}\rightarrow u_L W^{\mu}u_L^{\dagger} -
\frac{i}{g}\partial^{\mu}u_L u^{\dagger}_L \ ,
 \qquad B^{\mu} \rightarrow
u_Y B^{\mu}u_Y^{\dagger} - \frac{i}{g^{\prime}} \partial^{\mu}u_Y
u^{\dagger}_Y \ ,  \label{rotation} \end{equation} where $g$ and
$g^{\prime}$ are the standard electroweak coupling constants,
$W_{\mu}=W_{\mu}^a \frac{\tau^a}{2}$ and
$B_{\mu}=B_{\mu}\frac{\tau^3}{2}$.

A convenient method of coupling the electroweak gauge fields to the
globally invariant effective Lagrangian in Section \ref{ELSUSU} is
to introduce a covariant derivative, which includes the $W$ and $B$
fields as well as the strong vector and axial-vector fields,
\begin{equation} {\rm
D}^{\mu}M=\partial^{\mu}M - ig W^{\mu}M + i g^{\prime}MB^{\mu}
-i\tilde{
g}c C^{\mu}_L M +i \tilde{g}c^{\prime}M C^{\mu}_R \ ,
\label{newD}
\end{equation} where we have defined the new vector fields
$ C^{\mu}_L=A_L^{\mu} -\frac{g}{\tilde{g}}W^{\mu}$,
$C^{\mu}_R=A_R^{\mu} -\frac{g^{\prime}}{\tilde{g}}B^{\mu}
\label{cdef}$ and where $c$ and $c^{\prime}$ are arbitrary real constants (for
details see Ref.~\cite{ARS}).
 By requiring invariance under the parity operation exchanging
the labels $L\leftrightarrow R$ we have the extra condition
$c=c^{\prime}$.

The effective Lagrangian, constructed to be invariant under a local
$SU_{L}(2) \times U_{Y}(1)$ as well as $CP$ is through dimension
four,
\begin{eqnarray}
L&=&\frac{1}{2}{\rm Tr}\left[ {\rm D}_{\mu }M{\rm D}^{\mu
}M^{\dagger }
\right]+m^{2}\,{\rm Tr}\left[ C_{L\mu }C_{L}^{\mu }+C_{R\mu }C_{R}^{\mu }
\right] +h\,{\rm Tr}\left[ C_{L\mu }MC_{R}^{\mu }M^{\dagger
}\right] \nonumber \\ &+&{r}\,{\rm Tr}
\left[ C_{L\mu }C_{L}^{\mu }MM^{\dagger }+
C_{R\mu }C_{R}^{\mu }M^{\dagger }M
\right] + i\,\frac{s}{2}\,{\rm Tr}\left[ C_{L\mu }\left( M{\rm D}^{\mu
}M^{\dagger }-{\rm D}^{\mu }MM^{\dagger }\right) \right. \nonumber
\\
 &+& \left. C_{R\mu }\left(
M^{\dagger }{\rm D}
^{\mu }M-{\rm D}^{\mu }M^{\dagger }M\right) \right] .
\label{newLagrangian}
\end{eqnarray}
To this we add a kinetic term
\begin{equation}
L_{{\rm Kin}}= -\frac{1}{2}{\rm Tr}\left[ F_{L\mu \nu }F_{L}^{\mu
\nu
}+F_{R\mu \nu }F_{R}^{\mu \nu }\right] -\frac{1}{2}{\rm Tr}\left[
W_{\mu
\nu }W^{\mu \nu } \right] -\frac{1}{2}{\rm Tr}\left[ B_{\mu \nu }B^{\mu
\nu }
\right] \ ,
\label{generalkinetic}
\end{equation}
where $W_{\mu \nu } =\partial _{\mu }W_{\nu }-\partial _{\nu
}W_{\mu }-ig\left[ W_{\mu },W_{\nu }\right]$ and $B_{\mu \nu }
=\partial _{\mu }B_{\nu }-\partial _{\nu }B_{\mu }$.
One should still add other interaction terms involving the
$C_{L/R}$ fields,  the interaction term $\displaystyle{{\rm
Tr}\left[M M^{\dagger}\right]{\rm Tr}\left[C_L^2 + C_R^2
\right]} \label{Ctracedouble}$ and a symmetry breaking potential.

The extension of this effective Lagrangian to the relevant case of
the larger symmetry group $SU_L(N_f)\times SU_R(N_f)$ with $N_f >
2$, is straightforward.

The vector and axial vector masses computed in Ref.~\cite{ARS},
$M_V^2$ and $M_A^2$, are arbitrary, depending on the choice of
parameters, although generically we expect them and the scalar
masses to be of order $4\pi^{2}v^2$.

Expanding the Lagrangian up to quadratic terms in the fields one
observes the presence of weak mixing terms providing a contribution
from physics beyond the standard model to the oblique electroweak
corrections. (One can find the actual computations and more
explicit formulae in Ref.~{\cite{ARS}}). These may be described by
the $S$, $T$, and $U$ parameters, but the last two vanish in the
present model because there is no breaking of weak isospin in the
strong sector. The $S$ parameter receives contributions from all
the physics beyond the standard model, including, in the model
being used here, loops of pseudo-Goldstone bosons (PGB's), the
strongly interacting massive scalars, and the vector and axial
vector. The direct, vector-dominance contribution of the vector and
axial vector is
\begin{equation} S_{vect-dom} = \frac{8\pi}{\tilde{g}^{2}}\left[
\frac{M_{A}^{2}
\left( 1-\chi \right) ^{2}}{M_{Z}^{2}-M_{A}^{2}}-
\frac{M_{V}^{2}}{M_{Z}^{2}-M_{V}^{2}}\right]
 \approx \frac{8\pi}{\tilde{g}^{2}}  \left[1 - \left(1 -
 \chi\right)^2\right] .
\label{s-ciao}
\end{equation}
with $\displaystyle{\chi=\frac{v^{2}}{2 M_{A}^{2}}\tilde{g}\left[
\tilde{g}c-s\right]}$.
 Clearly, this contribution to the $S$
parameter can take on any value depending on the choice of
parameters. Its typical order of magnitude, with the strong
coupling estimate $\tilde{g}^2
\approx 4
\pi^2$, is expected to be $O(1)$.

The parameter choice $s=\tilde{g}c$ and $h=-\tilde{g}^{2}c^{2}$
gives $\chi = 0$, leads to the degeneracy of the vector and axial
vector and the vanishing of $S_{vect-dom}$. The further choice
$\displaystyle{r=\frac{\tilde{g}^2c^2}{2}}$ leads to the collapse
of the general effective Lagrangian into the simple form
\begin{equation} L=\frac{1}{2}{\rm Tr}\left[{D}_{\mu }M{D}^{\mu
}M^{\dagger}\right] +m^{2}\, {\rm Tr}\left[ C_{L\mu }C_{L}^{\mu
}+C_{R\mu }C_{R}^{\mu }\right] \ , \label{symmetric}
\end{equation}
along with the kinetic terms of Eq.~(\ref{generalkinetic}),
interactions among the $C_{L/R}^\mu$ fields and a symmetry breaking
potential. Here, $\displaystyle{DM=\partial M-igWM+ig^{\prime}MB}$
is the standard electroweak covariant derivative.

The strongly interacting sector has split into two subsectors,
communicating only through the electroweak interactions. One
subsector consists of the Goldstone bosons together with their
massive scalar partners. The other consists of the degenerate
vector and axial vector described by the $A_{L/R}^{\mu }$ fields.
In the absence of electroweak interactions, there is an enhanced
symmetry $\left[ SU_{L}(2)\times SU_{R}(2)\right] \times
\left[ SU_{L}(2)\times SU_{R}(2)\right]$, breaking spontaneously
to $ SU_{V}(2)\times \left[ SU_{L}(2)\times SU_{R}(2)\right]$. The
electroweak interactions explicitly break the enhanced symmetry to
$SU_{L}(2) \times U_Y(1)$. All of this may be generalized to $N_f >
2$, necessary to yield a near-critical theory.

The additional symmetry has an important effect on the $S$
parameter, suppressing contributions that are typically large in
QCD-like theories.

Finally, it could be that only a lesser, discrete symmetry emerges
in the physical spectrum. Even this would be sufficient to insure
vector-axial degeneracy and the vanishing of the vector dominance
contribution to the $S$ parameter.

\section{Conclusions}
\label{conc}

To explore some features that might arise in a strongly coupled
gauge theory when the number of fermions $N_f$ is near a critical
value for the transition to chiral symmetry an effective Lagrangian
approach was used.

The spectrum was taken to consist of a set of Goldstone particles,
associated massive scalars, and a set of massive vectors and axial
vectors. It was observed that parity doubling is associated with
the appearance of an enhanced global symmetry, consisting of the
spontaneously broken chiral symmetry of the underlying theory
($SU_{L}(N_f)
\times SU_{R}(N_f)$) together with an additional, unbroken
symmetry, either continuous or discrete. The additional symmetry
leads to the degeneracy of the vector and axial vector, and to
their stability with respect to decay into the Goldstone bosons.

Despite the hints in Refs. \cite{AS,mawhinney} it has not been
established that an underlying gauge theory leads to these enhanced
symmetries as $N_f$ approaches a critical value for the chiral
transition. If it is to happen an interesting interplay between
confinement and chiral symmetry breaking would have to develop at
the transition.

Finally by electroweak gauging a subgroup of the chiral flavor
group it was shown that the enhanced symmetry provides a partial
custodial symmetry for the $S$ parameter, in that there is no
contribution from the parity-doubled sector by itself.

\section*{Acknowledgments}

I am happy to thank Thomas Appelquist and Paulo Sergio Rodrigues Da
Silva for sharing the work on which this talk is based.
 The work of F.S. has been partially supported by the US DOE
under contract DE-FG-02-92ER-40704.

\end{document}